**Highlights:**

1. To the best of our knowledge, Brain Deep Embedded Cluster(BDEC) is the first study that uses deep learning algorithm for rs-fMRI-based cortical parcellation.
2. The BDEC parcellation considers global similarity, inter-class difference and spatial connectivity.
3. The structure of BDEC model is simple and easy to implement.
4. By extensively comparing with nine commonly used rs-fMRI-based parcellations, the BDEC parcellation performs better in terms of functional homogeneity, validity, network analysis, task homogeneity and generalization capability.

# BDEC:Brain Deep Embedded Clustering model

Xiaoxiao Ma[1,2], Chunzhi Yi[1,2,*], Zhicai Zhong[1,2], Hui Zhou[1,2], Baichun Wei[2], Haiqi Zhu[2], Feng Jiang[1,2,*]

[1]Zhengzhou Research Institute of Harbin Institute of Technology, Zhengzhou, Henan, China

[2]School of Medicine and Health, Harbin Institute of Technology, Harbin，Heilongjiang, China

[*]Corresponding authors: Email: chunzhiyi@hit.edu.cn, https://orcid.org/0000-0002-4180-1109; https://orcid.org/0000-0001-8342-1211

**Abstract**

An essential premise for neuroscience brain network analysis is the successful segmentation of the cerebral cortex into functionally homogeneous regions. Resting-state functional magnetic resonance imaging (rs-fMRI), capturing the spontaneous activities of the brain, provides the potential for cortical parcellation. Previous parcellation methods can be roughly categorized into three groups, mainly employing either local gradient, global similarity, or a combination of both. The traditional clustering algorithms, such as "K-means" and "Spectral clustering" may affect the reproducibility or the biological interpretation of parcellations; The region growing-based methods influence the expression of functional homogeneity in the brain at a large scale; The parcellation method based on probabilistic graph models inevitably introduce model assumption biases. In this work, we develop an assumption-free model called as BDEC, which leverages the robust data fitting capability of deep learning. To the best of our knowledge, this is the first study that uses deep learning algorithm for rs-fMRI-based parcellation. By comparing with nine commonly used brain parcellation methods, the BDEC model demonstrates significantly superior performance in various functional homogeneity indicators. Furthermore, it exhibits favorable results in terms of validity, network analysis, task homogeneity, and generalization capability. These results suggest that the BDEC parcellation captures the functional characteristics of the brain and holds promise for future voxel-wise brain network analysis in the dimensionality reduction of fMRI data.

***Keywords:*** *Resting state fMRI, cortical parcellation, deep learning*

# 1. Introduction

Functional magnetic resonance image (fMRI) captures the functional fluctuations of brain, thus emergingly contributes to large-scale brain network analysis. It may aid the understanding of the dynamics of brain activities and their correspondence with cognition, personalities and diseases. Such brain network analysis mainly relies on functional connectivity among cortical areas that measures the functional synchronization between cortical areas. The high dimensionality of the spatio-temporal fMRI images would pose a significant challenge for the storage and computing resources, if voxel-wise connectivity is used. According to the Segregation and Integration principle (Tononi et al., 1994), voxels can be averaged within a cortical area to represent the activity of the compact groups of neurons. In this way, the signal-to-noise ratio can be improved and the functional unit or the node of the brain network can be depicted. Thus, a rational parcellation and a functionally accurate labeling of cortical areas are of fundamental importance.

Among all the modalities for parcellation, resting-state fMRI (rs-fMRI) that captures the spontaneous activities of brain gradually becomes one of the top candidates. On one hand, compared with the anatomy-based methods that utilize anatomic markers (e.g. AAL) or cellular architecture information (e.g. Brodmann), rs-fMRI-based methods consider the functional fluctuations of brain can provide a more rationale parcellation for network analysis. Anatomy-based methods derive from the idea that cortical areas with different topographies and architectures correspond to different functions. However, there is no evidence of a one-to-one correlation between brain structures and functions. Parcellating cortical areas in the sense of neuro-genetics for functional network analysis is not particularly rationale. On the other hand, compared with other non-invasive brain imaging techniques, rs-fMRI-based parcellation would benefit the connectivity studies of fMRI. rs-fMRI-based parcellation considers the rationally functional fluctuations and cluster voxels with similar activities, can thus provide an effective dimension reduction for fMRI. Other than the benefit of being functionally related, studies presented that rs-fMRI carried sufficient information of anatomical constraints (Biswal et al., 1995) and correlations with gene expressions. Consequently, rs-fMRI-based parcellation increasingly becomes a popular manner of analyzing functional connectivity and large-scale brain networks.

The rs-fMRI-based parcellation methods identify cortical areas using data-driven approaches and considering the information of either local gradient or global similarity. One family of such approaches is clustering algorithms. Algorithms like K-means, spectral clustering, hierarchical clustering, and mixture models were used to cluster voxels with similar activities. Such methods mainly consider the global similarity of brain activities and treat the voxels with similar activities as a function unit, attenuating or even ignoring the spatial locations and contiguity of cortical areas. Furthermore, such traditional clustering algorithm or mixture model-based methods are sensitive to initialization or nearly equal-size parcels thus may affect the reproducibility or the biological interpretation of parcellations.

Another family of parcellation approaches is mainly based on region growing (Gordon et al., 2016). Such approaches rely on the assumption of local gradients. That is, different adjacent parcels should present an abrupt change of functional connectivity(FC) which can be measured by the significantly large local gradients of FC. Region growing-based methods thus can delineate the function units in a spatially continuous manner and encode the geometry of the cortex similar to the histological parcellation of cortex (Wig et al., 2014). In this way, the parcellation may fit the biological boundary well. However, compared with clustering methods that utilize global similarity, local gradient constrains the spatial congruity and may affect the homogeneity of brain activities within a parcel. Moreover, the region growing approaches dependent on seed regions still suffer from the sensitivity of initial seed selection.

Following studies integrate the information of local gradient and global similarity under certain model biases. Thirion concatenated spatial coordinates with rs-fMRI and used K-Means for clustering (Thirion et al., 2014). Baldassano et al. posed a normal distribution on connections and used a non-parametric Bayesian model to achieve considerable region connectivity without worrying about initialization sensitivity (Baldassano et al., 2015). Honnorat realized an initialization-free parcellation by modelling the generation of rs-fMRI with Markov Random Field (MRF) and adding a shape prior on the connectedness of parcels (Honnorat et al., 2015). Schaefer used gradient-weighted MRF to segment the cortex and added spatial constraints, resulting in 100-1000 parcels (Schaefer et al., 2018). Although both local gradient and global similarity are integrated to achieve a trade-off between neurobiological interpretation and signal homogeneity, the models adopted by previous work are constrained by certain model biases, such as prior assumption, Bayesian assumption or Markov assumption. Such biases may induce an implicit bias of final parcellations.

In this work, we develop an assumption-free model that utilizes the strong data fitting ability of deep learning. We design a loss function that considers signal homogeneity within parcels, signal heterogeneity among parcels and spatial congruity among parcels. To the best of our knowledge, this is the first study that uses deep learning algorithm for rs-fMRI-based parcellation. The key idea of the work is to use inter-class distance to maximize the signal difference among parcels, use intra-class loss to maximize the signal similarity within parcels and encode spatial coordinates of voxels into feature vectors to encourage spatial congruity. A deep clustering method is adapted. Extensive experiments are conducted to compare our model with nine other parcellations. Even under the same number of parcels, our model presents a higher functional homogeneity, task homogeneity, validity and benefits for network analysis.

## 2. Materials and Methods

### 2.1. Overview

The HCP S1200 release dataset is randomly divided equally into a training set (N=548)

and a test set (N=548). A deep learning based parcellation is developed and applied to the train set. The resulting parcellations are compared with nine previously published rs-fMRI parcellations using multimodal data from multiple protocols with diverse acquisition and processing protocols.

## 2.2. HCP Dataset

We employed a large dataset from the HCP S1200 release, comprising almost all available data from 1096 subjects aged between 22 and 35 years. High-resolution T1w (TI=1000ms, TR=2400ms, TE=2.14ms, FA=8 degrees, FOV=224mm, matrix=320, 256 sagittal slices) and T2w images (TR=3200ms, TE=565ms, FOV=224mm, matrix=320) were collected by the HCP. The T1w and T2w data underwent preprocessing using a custom pipeline developed by HCP and FreeSurfer to generate detailed surface meshes of white/gray matter, as well as the brain surface/cerebrospinal fluid interface. These meshes were spatially normalized to Montreal Neurological Institute (MNI) space and resampled to 32k vertices as fs_LR_32k.

The authors named these acquisitions REST1_LR, REST1_RL, REST2_LR and REST2_RL, with the images obtained on two separate days: REST1_LR/REST1_RL on day one and REST2_LR/REST2_RL on another day. Each session included 1,200 time points and the BOLD session length was approximately 15 minutes. The dataset was preprocessed and denoised using the HCP structural and functional minimal preprocessing pipelines (Glasser et al., 2016), with no additional global signal regression, tissue regression, temporal filtering, or motion scrubbing.

The HCP S1200 dataset is divided into training (N=548) and test (N=548) sets using only REST1 data, and the data from each set was subsequently averaged to obtain an rs-fMRI data file representing the group level. The model is trained on the training set, and all comparative experiments are conducted on the test set.

## 2.3. Parcellation Methods

The cortex parcellation problem is essentially a clustering problem. In recent years, there has been an interest in using deep learning models to perform clustering, which we will call "deep clustering" for short. Deep clustering is a model that uses neural networks combined with clustering algorithms for unsupervised learning. It performs clustering tasks by learning low-dimensional representations of high-dimensional data, which can automatically learn features with semantic meaning from the original data and map the data points to the low-dimensional space for clustering. Herein, following the protocol presented in (Xie et al., 2016), the deep clustering model consists of an encoder-decoder architecture and a clustering algorithms. The encoder maps the raw data into a low-dimensional space and produces encoded vectors as input to the clustering algorithm. Then, the encoded vectors can be clustered accordingly.

We propose a deep clustering model for the task of brain parceling and experimentally demonstrate that the parcellations generated by this model outperform most of the existing parcellations. Figure 1 shows the structure diagram of the whole model and the process diagram of the parallelization generation. To be specific, the encoder is firstly used to yield a low-dimensional representation of the signals of each brain vertex. The location of each brain vertex is softly embedded into the encoded vectors. In this way, the vectors used for consequent clustering carry the information of both functional fluctuations and the geometrical characteristics of the brain vertices. After clustering the vectors by K-means++, we calculate the similarities among vectors and their corresponded cluster centers and further enlarge the similarities by projecting them on a designed distribution. Then, the distance among cluster centers are used to increase inter-class distance. Finally, the reconstruction error, intra-class and inter-class losses are integrated to train the whole architecture. The loss function can be formulated as

$$L = L_{res} + \alpha L_{clu} + \beta L_{dis} \tag{1}$$

where $L_{res}$ denotes the reconstruction error of the encoder-decoder architecture, $L_{clu}$ denotes the loss of intra-class distance enlarged by the distribution projection, $L_{dis}$ denotes the loss of inter-class distance, $\alpha$ and $\beta$ are both taken as 0.01 in this model.

$$L_{res} = \frac{1}{N}\sum_{i=1}^{N}\| x_i - \hat{x}_i \|^2 = \frac{1}{N}\| X - \hat{X} \|^2 \tag{2}$$

where $N$ denotes the number of the samples, i.e., the number of the brain vertices. By using the reconstruction loss, the latent and compact representation can be obtained, which are usually demonstrated with refined information compared with solely using signals(Ma et al., 2019; Zhou et al., 2018). The information refining ability enabled by the deep learning-based encoder should be especially noted, especially considering that previous work on rs-fMRI-based parcellation directly utilizes the signals of each brain vertex for clustering.

The Intra-class loss $L_{clu}$ and the inter-class loss $L_{dis}$ correspond to the global similarity term that assigns the same label to the brain vertices with similar signal representations and the interregional heterogeneity term that tends to maximize signal differences in each parcel to capture of the different functional structures of the brain. The intra-class loss $L_{clu}$ is given by

$$L_{clu} = KL(P \| Q) = \sum_i \sum_j p_{ij} \log \frac{p_{ij}}{q_{ij}} \tag{3}$$

where $q_{ij}$ and $p_{ij}$ denote the distributions of cluster centers and the projected distribution, respectively, $Q$ and $P$ denote the distribution consisted of $q_{ij}$ and $p_{ij}$, respectively. As what will be shown in section, $q_{ij}$ measures the distance among the vectors and their cluster centers and $p_{ij}$ enlarges $q_{ij}$. By minimizing $L_{clu}$, a compact cluster can be expected.

$$L_{dis} = \frac{M*M}{\sum_i^M \sum_j^M \| u_i - u_j \|^2} \tag{4}$$

where $M$ denotes the number of clusters, $u_i$ and $u_j$ denote the $i-th$ and $j-th$ cluster center. By minimizing $L_{dis}$, a larger distance among clusters can be expected and the representations of clusters can be more distinguishable.

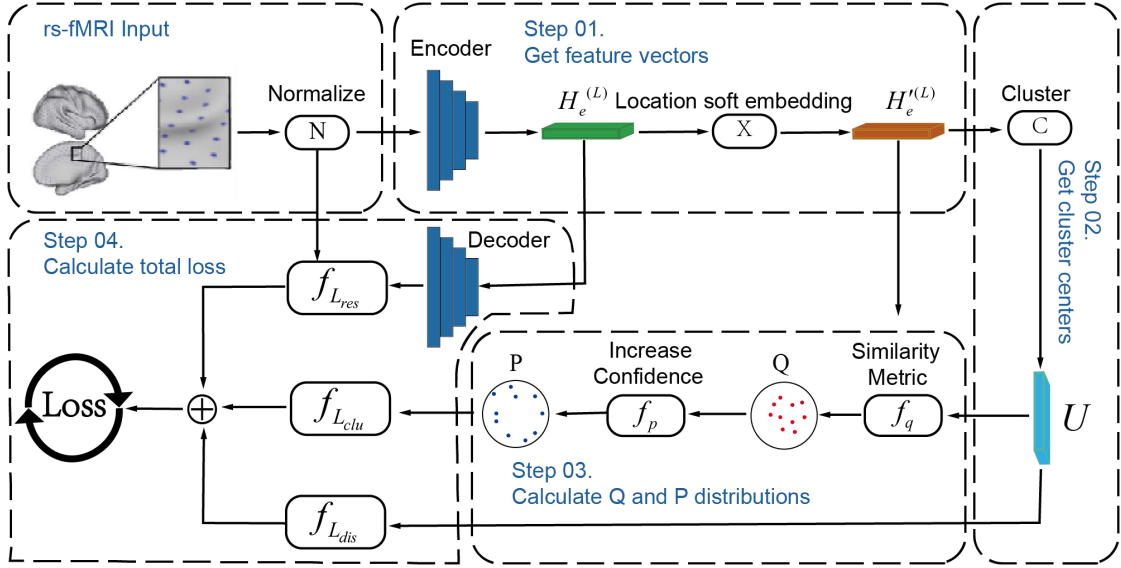

**Figure 1.** An overall flowchart of the cerebral cortex parcellation process.

**1) Signal encoding and location embedding:** Due to the high dimensionality of the time series of rs-fMRI data, dimensionality reduction is required. An autoencoder is an unsupervised learning model commonly used for dimensionality reduction and feature extraction. By learning a low-dimensional representation of the input data, autoencoders can map high-dimensional data to a low-dimensional space while retaining the information of the original data, so we choose autoencoders as a tool for data dimensionality reduction. Specifically, the representation learned by the $i-th$ layer, in encoder part, $H_e^{(l)}$, can be obtained as follows.

$$H_e^{(l)} = \phi(W_e^{(l)} H_e^{(l-1)} + b_e^{(l)}) \tag{5}$$

where $\phi$ is the activation function of the fully connected layers such as Relu function, $W_e^{(l)}$ and $b_e^{(l)}$ are the weight matrix and bias of the $l-th$ layer in the encoder, respectively. Besides, we denote $H_e^{(0)}$ as the raw data, which is rs-fMRI data.

We perform location soft embedding on $H_e^{(L)}$, and get embedded representation called $H_e'^{(L)}$, given by:

$$h_i' = h_i \times \cos x_i \times \cos y_i \times \cos z_i \tag{6}$$

where $h_i$ and $h_i'$ is the $i-th$ sample of $H_e^{(L)}$ and $H_e'^{(L)}$, respectively. The

$\cos x_i$, $\cos y_i$, and $\cos z_i$ represent the cosine of the three-dimensional coordinates of vertex $i$, respectively. We consider that the location embedding of neural networks does not require particularly elaborate mathematical derivation. For example, the transformer model, used trigonometric position embedding, which is mathematically detailed. Later, a larger transformer model, called Bert, used learnable location embedding without any mathematical derivation. Through comparison, it is found that the effect of the model is not worse.(Devlin et al., 2019; Vaswani et al., 2017)

**2) Clustering:** We perform the deep clustering iteratively. A traditional clustering algorithm is used to determine the initial cluster center. We use K-means++ to cluster $h_i'$ and get the initial cluster center $u_i$. In subsequent experimental analyses, we evaluate the performance under different numbers of clusters and determine the cluster number as 400.

**3) Distribution projection:** When calculating the intra-class loss $L_{clu}$, the distance of the vectors and their corresponded cluster centers is firstly calculated using the Student's t-distribution as a kernel which is a commonly used kernel in the field of image depth clustering.

$$q_{ij} = \frac{(1+\|h_i' - u_j\|^2 / t)^{-\frac{t+1}{2}}}{\sum_{j'}(1+\|h_i' - u_{j'}\|^2 / t)^{-\frac{t+1}{2}}} \tag{7}$$

where $u_j$ denotes the $j-th$ cluster center, $t$ denotes the degree of freedom of the Student's t-distribution. In this work we only set $t=1$, $q_{ij}$ denotes the probability of assigning sample $i$ to cluster $j$, i.e., a soft assignment. We aim to further optimize the data representation by learning from the high confidence assignments. Specifically, a target distribution $P$ is designed to make data representation closer to cluster centers, thus improving the cluster cohesion.

$$p_{ij} = \frac{q^2_{ij}/f_j}{\sum_{j'} q^2_{ij'}/f_{j'}}$$

$$f_j = m_j + v_j + c$$

$$m_j = \sum_i q_{ij} \qquad (8)$$

$$v_j = \sum_i \sum_j \sqrt{-\frac{N_j}{\sum_k N_k(1-q_{ij})^2 \log(q_{ij})}}$$

where $m_j$ denotes the soft cluster frequency, i.e., the summed probability of the vectors assigned to the $j-th$ cluster. $N_j$ denotes the number of vectors assigned to the $j-th$ cluster, $v_j$ utilizes $N_j$ to encourage the enlargement of small areas, $f_j$ blends information from $m_j$ and $v_j$ in order to further integrate both the summed probabilities and number of vectors within a cluster, $C$ is a vital hyperparameter to control the relative magnitude of $m_j$ and $v_j$. In this way, each $q_{ij}$ is squared and normalized so that the assignments can have higher confidence.

We denote the distributions consisted of $p_{ij}$ and $q_{ij}$ by $P$ and $Q$, respectively. By minimizing the KL divergence loss between $P$ and $Q$ distributions, the target distribution $P$ can help the model learn a better representation for clustering task, i.e., enabling a compact data representation inside a cluster. This is regarded as a self-supervised mechanism, because the target distribution $P$ is calculated by the distribution $Q$, and the $P$ distribution supervises the updating of the distribution $Q$ in turn. In this way, the total loss can be calculated and used to train the architecture to obtain the representations and assignments of each brain vertices.

### 2.4. Training-processing

Our encoder is designed with dimensions of $d$-500-500-2000-10, where $d$ is the dimensionality of the input data, which is 1200 in HCP S1200 release. Decoder is the symmetric structure of encoder. We first train the autoencoder with iterative decoding and encoding for 500,000 iterations. We then use the pre-trained autoencoder to train our model, where the model is iteratively trained for 200,000 times. We use a learning rate starting with $1e^{-3}$ and warm up 10,000 iterations in advance. The learning rate is gradually lowered by multiplying it with the cosine of the ratio between current iteration and the maximum iterations we allow. The minimum learning rate is set as $1e^{-5}$. We also set batchsize to 1 and use Relu as the activation function.

Considering that the drastic changes of the target distribution $P$ between adjacent iterations may induce the divergency issue, we adopt the following formula to slow down the changes in $P$ between iterations:

$$P = 0.1 \times P + 0.9 \times P_{pre} \qquad (9)$$

where $P_{pre}$ is the previous iteration of $P$.

### 2.5. Post-processing

Considering the inherent noise of rs-fMRI signals and the soft location embedding we adopt, the parcellation obtained above may be scattered. It is desirable to parcellate adjacent vertices into the same region to alleviate these unwanted effects. Herein, we define a parcel as a collection of vertices with the same label value and a region if the vertices are still spatially connected. A parcel can be composed of multiple regions. In order to further contain the number of regions within a parcel, we propose two post-processing methods to improve the regional connectivity, i.e. the first one called "removed," and the second one called "merged."

For the "removed" method, we remove regions with fewer than 20/9 vertices on left/right hemisphere separately (considering the size and number of parcels) and then dilate the entire parcellation to fill any gaps. Since we only remove regions, a parcel can still correspond to multiple regions, but the number of regions is greatly reduced. Therefore, the post-processing method is a mild post-processing.

For the "merged" method, our approach is to merge small regions into nearby regions with the highest regional homogeneity and based on "remove" method. The entire algorithm is shown in Algorithm 1. We first identify all parcels that contain more than one region and save them in a list. Then we find the smallest area region, calculate its regional homogeneity with all nearby regions and further merge it into the nearby region with the highest regional homogeneity. We iterate the above process until the list is empty.

**Algorithm 1** Merge post-process

**Require:** The Original parcellation label, $L_{ori}$; The average rs-fMRI data, $DATA$.

**Ensure:** The Merged parcellation label, $L_{merged}$.

1: Using the depth-first search algorithm to find all disconnected regions from $L_{ori}$, called $L_{reg}$.

2: Finding all parcels for more than one region and store all regions index to a list, $List$.

3: **while** List is not empty **do**

4:      Getting the region of smallest size, $Reg_{min}$.

5:      Using DATA to calculate the correlation between $Reg_{min}$ and it's connected neighbor regions.

6:      Choose the pair regions which have the max correlation and merge them in $L_{ori}$.

7:      Removing $Reg_{min}$ from List.

8: **end while**

9: Setting $L_{merged} = L_{ori}$.

10: Return $L_{merged}$.

## 2.6. Evaluation Metrics

If the parcellation indeed reflects how the cerebral cortex is working and organizing as a functional organ, each parcel should have high functional homogeneity and the network of regions should better reflect the brain organization. In this way, to what extent our parcellation corresponds to the spontaneous and task-evoked organizations of the brain can be tested. We consider evaluation metrics based on function, task, network performance. Under such evaluation metrics, we can compare this model parcellation with other publicly available parcellations.

**2.6.1 Function homogeneous**

It is important for a parcel to exhibit a consistent and homogeneous connectivity pattern, indicating that the connectivity pattern within the parcel is uniform. Hence, the homogeneity of the created parcels can serve as a quality metric for the parcellation process, as noted by (Craddock et al., 2012; Shen et al., 2013). However, it should be noted that parcel homogeneity is likely to be affected by parcel size, with smaller parcels inherently having higher homogeneity. This phenomenon can be observed by dividing a large parcel into smaller ones, which can still exhibit a high degree of functional homogeneity(Schaefer et al., 2018). Furthermore, data preprocessing methods (e.g. smoothing) can also increase the functional homogeneity of smaller parcels. To mitigate the impact of such issues on experimental results, we utilize null models to show how the parcellation improves the homogeneity within each parcel compared with randomly parceling with the same number of parcels.

**NULL MODEL**(Gordon et al., 2016). To evaluate the quality of the parcellation, we construct a null model consisting of randomly placed parcels of the same size, shape, and relative position as the original parcellation. This involves rotating each hemisphere of the original parcellation randomly around x, y, and z axes on the spherical expansion of the fs_LR_32k cortical surface. We repeat this process 1000 times to generate distributions of average homogeneity calculated from randomly placed versions of each tested parcellation. Due to non-uniform vertex density across the surface of the sphere, each parcel was slightly dilated to adjust for vertices gained or lost. It should be noted that some parcels were rotated into regions where no data existed (e.g., medial wall), and the homogeneity of these parcels cannot be calculated. An example of NULL MODEL is like Figure 2. We assigned them the average homogeneity of all random versions of the parcel that were rotated into valid cortical regions. The z-score was calculated as the difference between the original parcellation homogeneity and the distribution of random homogeneities, divided by the standard deviation of random homogeneities [(original homogeneity - mean of random homogeneities)/standard deviation of random homogeneities]. A higher z-score indicates a better parcellation quality and more consistent connectivity pattern within the parcels.

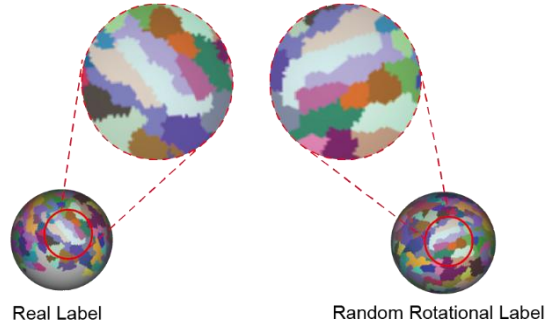

Figure 2. This is an example for NULL MODEL that the parcels with same color represent a parcel is placed randomly.

**DCBC**(Zhi et al., 2022). DCBC is an acronym for Distance-Controlled Boundary Coefficient. It is an unbiased assessment standard for the evaluation of discrete brain regions. It advances the field of functional brain mapping by comparing the predictive ability of different brain regions to define the most functionally different brain regions. DCBC can effectively eliminate the influence of data smoothing on the homogeneity of computing functions. In order to achieve this effect, it first divides the pairs of vertices within a certain distance into a bin. Second, the vertex pairs of the same bin are divided into two categories. One is within the same parcel (within), the other is not within the same parcel (between). The average Pearson correlation coefficients between vertices are calculated for each set (within or between) and averaged across vertex pairs of each set. The differences between the two sets for distance range $i$ are denoted as $p'_i$. The subscript $i$ represents the bin of the pairs of vertices corresponding to the distance $i$. Finally, by summing $p'_i$ of each distance weighted, the final DCBC value can be obtained. The formula is as follows:

$$DCBC = \sum_i w_i p'_i$$
$$w_i = \frac{n_{w,i} n_{b,i}}{n_{w,i} + n_{b,i}} / \sum_j \frac{n_{w,j} n_{b,j}}{n_{w,j} + n_{b,j}}$$

(10)

Where $n_{w,i}$ and $n_{b,i}$ represent the number of vertex pairs within and outside a distance range $i$, respectively.

**2.6.2 Task homogeneous**

Task-fMRI data from the Human Connectome Project (HCP) Group Average release in fs_LR_32k surface space was utilized to evaluate the task homogeneity of the parcels. This data set comprises 7 cognitive domains: social cognition, motor, gambling, working memory, language processing, emotional processing, and relational processing, which have 86 subdivided items. If the variance of task activation values within a region is smaller, it can indicate to some extent that the quality of parcel division is higher. Due to multiple tasks, we use the mean of each type task activation value variances within a parcel to represent task homogeneous for that parcel. We calculate 7 results for corresponding to each task. Considering that the variance decreases with the area of the regions, we also use null model for presenting a fair comparison.

**2.6.3 Validity**

Validity is also a key aspect in measuring the quality of parcellation, denoting the Silhouette coefficient (SC) is the most common index used to measure validity, and it indicates how well each vertex is assigned to its corresponding parcel. For each vertex, the SC computes the within-parcel distance, which is the average distance to all other vertices in the same parcel, and compares it to the inter-parcel distance, which is calculated from the distances between vertices assigned to different parcels. The SC not only evaluates the compactness of the parcels, but also the degree of separation between them. It is defined as follows:

$$SC_i = \frac{b_i - a_i}{\max(a_i, b_i)} \quad (11)$$

Given a parcellation $U = \{U_1, U_2, ... U_K\}$, $a_i$ and $b_i$ are defined as the within-parcel and inter-parcel distance of vertex $v_i \in U_k$, respectively the definitions of these measures are as follows:

$$a_i = \frac{1}{n_k - 1} \sum_{j \in U_k, i \neq j} d(v_i, v_j)$$
$$b_i = \frac{1}{M} \sum_{j \in \mathbb{N}(U_k)} d(v_i, v_j)$$
(12)

Here, $n_k$ denotes the number of vertices in $U_k$, $\mathbb{N}(U_k)$ denotes the parcels which are neighbors of $U_k$, with $M$ being the number of vertices within these neighboring parcels and $d(v_i, v_j)$ is the distance function, which is defined as $1-r$, where $r$ is Pearson's correlation computes between $v_i$ and $v_j$. We calculate inter-parcel distance through neighboring parcels rather than directly through all other parcels. This is because parceling methods usually assign highly homogeneous vertices to the same parcel. If we calculate inter-parcel distance through all other parcels, this will result in high SC values which are not conducive to comparison. Therefore, we use neighboring parcels to calculate inter-parcel distance to reduce SC values and make a fair comparison among parceling methods.

**2.6.4 Network analysis**

Parcellations are a valuable tool in reducing the dimensionality of the dense human connectome, while still retaining crucial information of interactions between different brain regions and the mechanisms that give rise to complex cognitive processes. The choice of parcellation method can have a significant impact on network analysis. We chose a task to explore how underlying parcellation affects network analysis: a network-based classification task.

One such classification task we explore is gender classification, as several studies have identified differences in both structural and functional connectivity between genders(Gong et al., 2011). Specifically, significant differences in the topological organization of functional networks have been found between males and females in terms of functional connectivity derived from rs-fMRI data(Tian et al., 2011). To evaluate the impact of the parcellation on this task, we use Gauss Support Vector Machine (SVM) (Burges, 1998), a well-established classifier, and a 10-fold cross-validation procedure to estimate each method's performance.

The goal of SVM is to find the hyperplane that separates the data into two classes with the largest margin. In other words, it aims to identify a (p-1)-dimensional hyperplane that represents the largest separation or margin between the feature vectors of the two classes.

With group-wise parcellations ensuring node correspondences, an embedding of each subject's connectivity matrix can be used to obtain a general vector representation(Varoquaux and Craddock, 2013) and assigned labels by the trained SVM. This approach is often referred to as "bag of edges" (Craddock et al., 2012) and has been widely used when the underlying parcellation is the same among all subjects.

**2.7. Selection of model hyperparameters**

For unsupervised clustering tasks, selecting an appropriate auxiliary clustering distribution $P$ is crucial. We found that the hyperparameter $C$ used to construct the $P$ distribution is critical to the clustering results, as $C$ is the denominator and controls the relative size of each element in $P$. Therefore, under the premise of each hemisphere having a resolution of 200 and with all other model hyperparameters fixed, we used the null model method to calculate functional homogeneity. By comparing the value of the z-score for different values of $C$ from 0-20, we determined the final value of $C$.

**2.8. Comparison with other Parcellations**

We compared the parcellation predicted by our Brain Deep Embedded Cluster (BDEC) model with nine other parcellations generated by different methods. These parcellations included: 'Baldassano' - a parcellation created by a nonparametric Bayesian model (Baldassano et al., 2015); 'Gordon' - a parcellation created by averaging gradients of resting-state functional connectivity networks (Gordon et al., 2016); 'Schaefer' - a parcellation created by a gradient-weighted Markov Random Field model (Schaefer et al., 2018); 'Fan' - a parcellation created using anatomical landmarks and connectivity-driven information (Fan et al., 2016); 'Shen' - a parcellation created using a spectral clustering approach (Shen et al., 2013); 'Glasser' - a parcellation created using a semi-automated approach on multimodal images of 210 adults (Glasser et al., 2016); 'JOINT' - a parcellation created by joint spectral decomposition of individual subjects (Arslan et al., 2015); 'K-Means-AVR' - a parcellation created by applying k-means clustering on concatenated functional connectivity matrices and spatial coordinates (Thirion et al., 2014); and 'GRASP' - a parcellation created by incorporating shape priors into a Markov Random Field (Honnorat et al., 2015). All of the parcellations mentioned above are projected into fs_LR_32k surface space, using the method provided by (Arslan et al., 2018), and their detailed descriptions are shown in Supplementary Table S1."

For the comparison of functional homogeneity, we used a full test set (N=548) and five-fold cross-validation, applied to both the null model approach (with 1000 random rotations) and the DCBC method. For the gender-based network analysis, we randomly selected 800 participants from the HCP S1200 release dataset, consisting of 400 males and 400 females. For the comparison of task homogeneity, we also used the test set (N=548) and five-fold cross-validation, applied to the null model approach.

## 3. Results

### 3.1. Confirm the resolution of parcellation

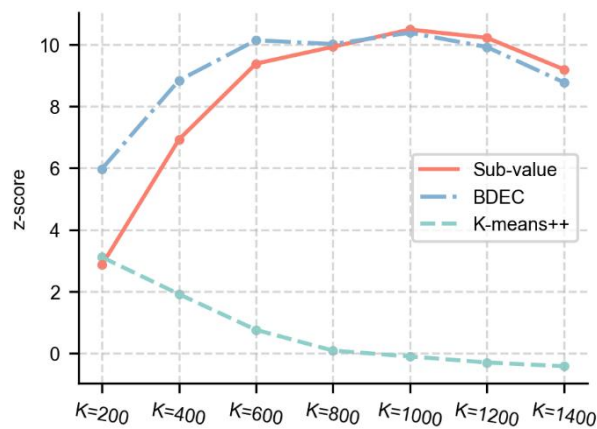

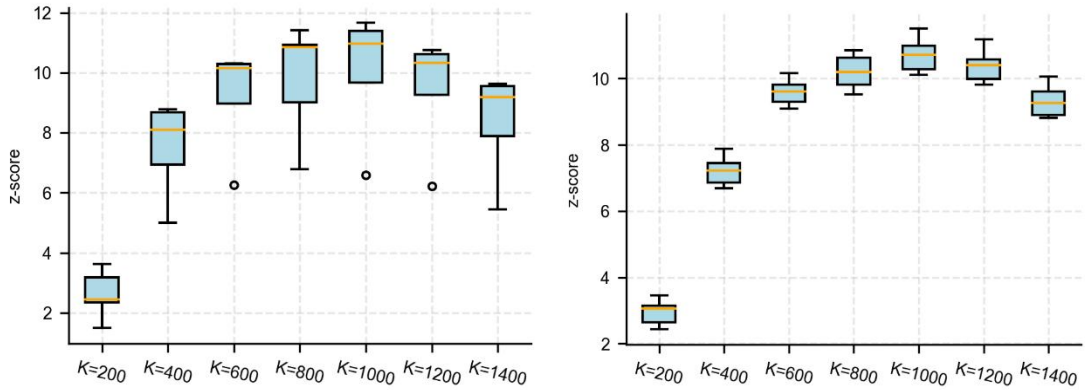

**Figure 3.** The comparison between BDEC and K-means++ in terms of their z-score at different parcel numbers K. **Top:** The z-score on the whole test set. **Bottom:** The difference between BDEC and K-means++ under five-fold cross-validation on the five validation sets (**Left**) and on the five sub-test sets (**Right**).

The Top of Figure 3(red line) presents the z-score difference between BDEC and K-means++ calculated by null model method of different parcellation resolutions. K-means++ was applied to $h_i'$, which is the feature vector that has been dimension-reduced and location soft encoded by the autoencoder. None of them used any post-processing. When the number of parcels reach 400, the increase of performance slows down. Additionally, it is also shown that BDEC presents a better performance than K-means++ irregular of parcel numbers. The fact is stable across datasets of different sizes, with the sub-test set being a large dataset (Bottom, Left) and the validation set being a small dataset (Bottom, Right).

Considering that the resolution of most publicly available parcellations is below 400 and the above fact, we ultimately chose 400 as the final resolution.

## 3.2. Selection of model hyperparameter $C$

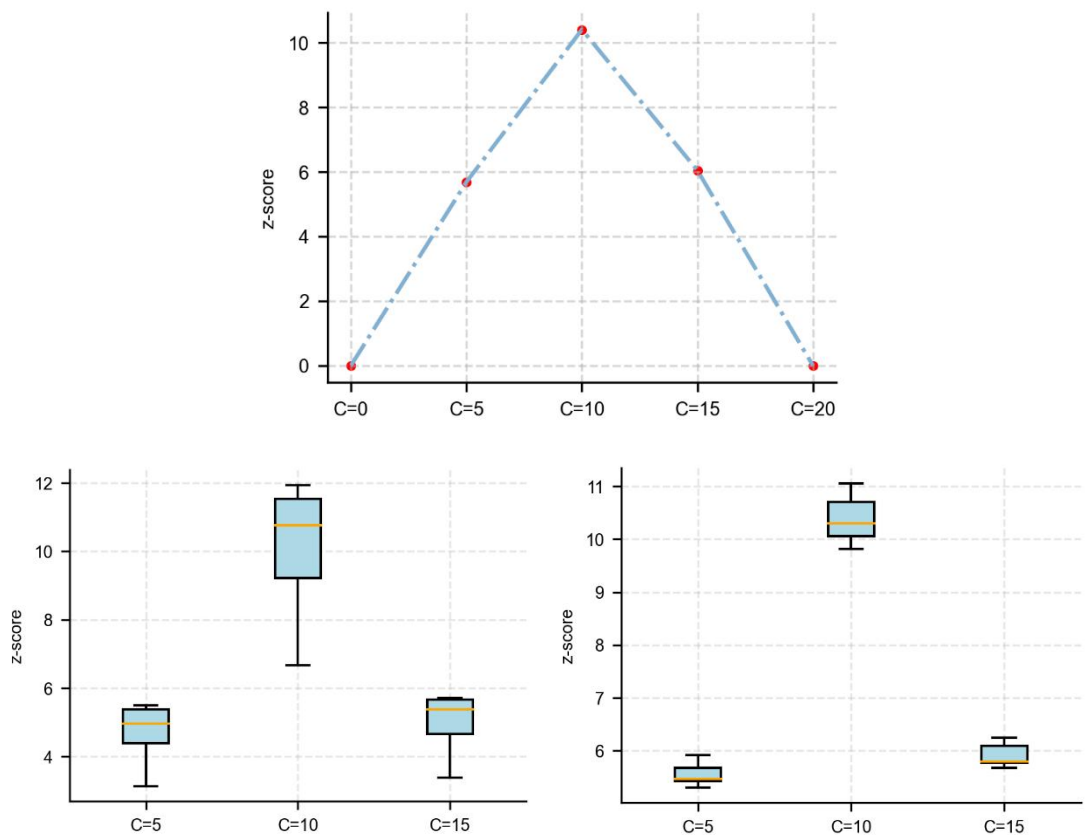

**Figure 4.** The optimal $C$ value by comparing z-score values with different $C$ values under the premise that K is 400. **Top:** The z-score on the whole test set. **Bottom:** The difference results of five-fold cross-validation on the whole test set, with the z-score on the five validation sets (**Left**) and on the five sub-test sets (**Right**).

The hyperparameter $C$ is important for constructing the auxiliary probability distribution $P$, which is a crucial distribution for guiding the unsupervised clustering model. We trained five models with different $C$ values of 0, 5, 10, 15, and 20 using the training set and a parcellation resolution of 400. We compared the parcel homogeneity of each model on the full test set using the null model method. As shown in the figure 4, the z-score is maximized when $C$ equals 10, and the change in z-score exhibits a trend that closely resembles a concave function. It is worth noting that when $C$ equals 0 or 20, there is a significant degradation in parcellation, which means that the final parcellation resolution is far lower than the specified value K. However, the experiment demonstrated that our choice of $C$ has strong interpretability.

## 3.3. Parcel Homogeneity with null model

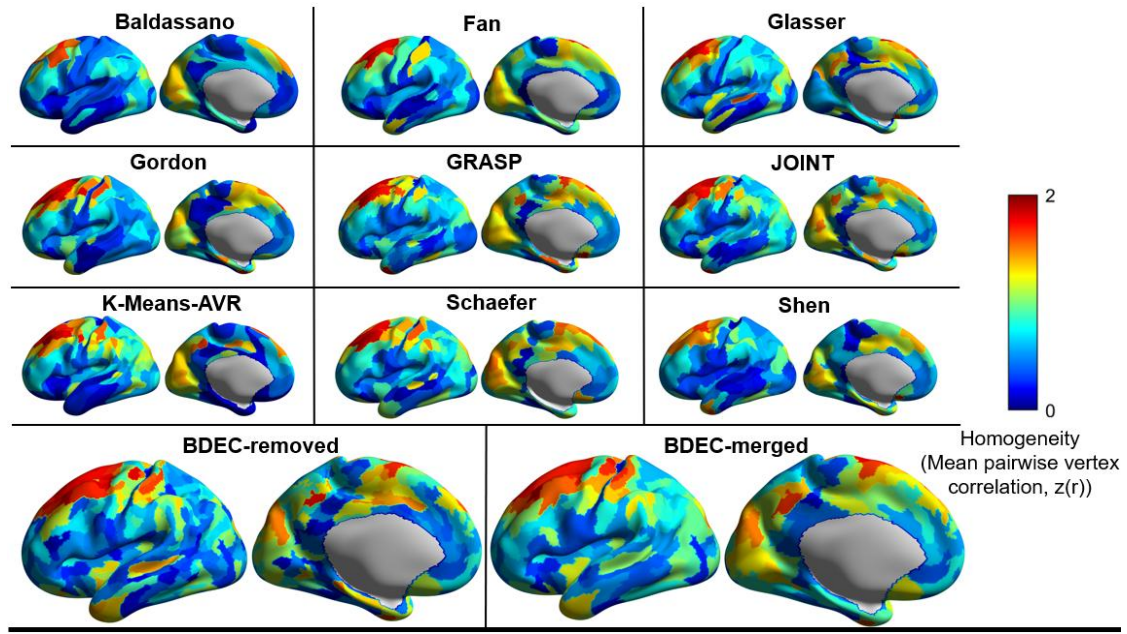

**Figure 5.** Using the average data of full test set to calculate the average Pearson correlation coefficient(r) within each parcel, and then apply the Fisher transformation to $r$. For the clarity of figure, we show only the left hemisphere.

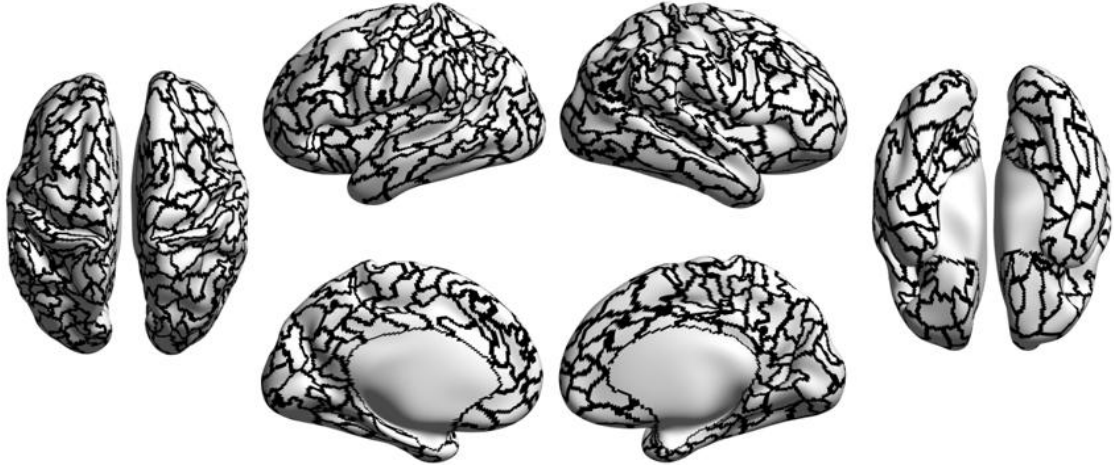

**Figure 6.** Boundary of BDEC-merged parcellation, left(K=198) and right(K=196) hemisphere.

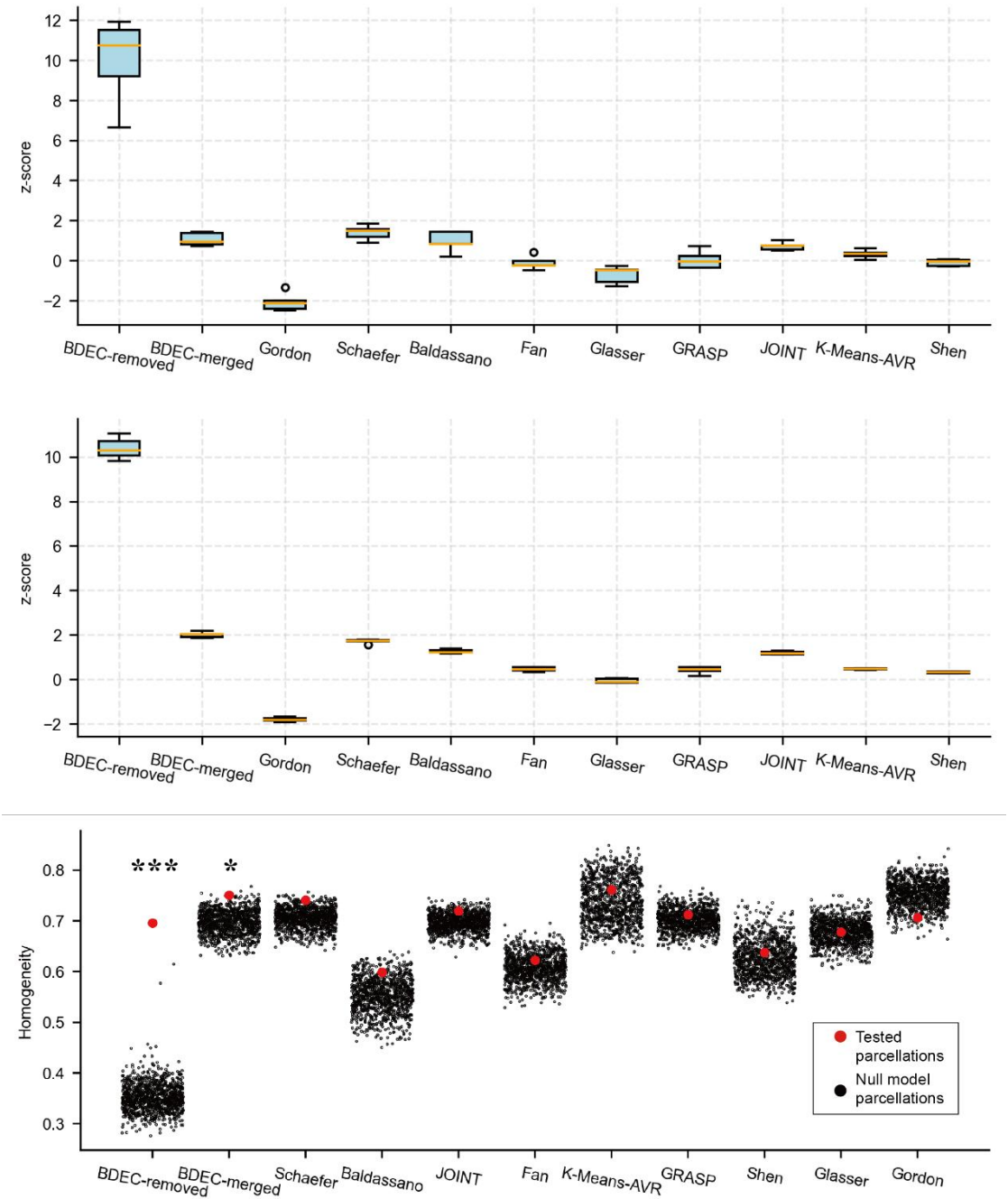

**Figure 7.** Compare the two BDEC parcellations with the other nine commonly used parcellations. It shows the difference results of five-fold cross-validation on the whole test set, with the z-score on the five validation sets (**Top**) and on the five sub-test sets (**Middle**). **Bottom:** It shows the homogeneity on the whole test set, average homogeneity across parcels of real parcellation (red dots) compared with the average homogeneity across parcels of each of 1000 null model iterations (black dots). *** indicates the parcellation was more homogenous than all of its 1000 null model iterations (i.e., $P < 0.001$). * indicates the parcellation was more homogenous than all of its 990 null model iterations (i.e., $P < 0.01$).

**Table 1.** It shows the z-score homogeneity on the whole test set.

| Parcellation | z-score |
| --- | --- |
| **BDEC-removed** | **10.4070** |
| **BDEC-merged** | **2.1897** |
| Schaefer400 | 1.7637 |
| Baldassano | 1.3017 |
| JOINT | 1.2725 |
| K-Means-AVR | 0.5567 |
| Fan | 0.5541 |
| GRASP | 0.4840 |
| Shen | 0.3761 |
| Glasser | 0.0192 |
| Gordon | -1.7527 |

From Figure 5, it can be visually observed that the homogeneity of our two BDEC parcellations is greater than other Parcellations. Compared to Baldassano, Fan, Gordon, and K-Means-AVR, the high homogeneity advantage of the BDEC parcellations is visible to the naked eye. Figure 6 shows the boundary of BDEC-merged with two hemispheres.

We used the null model method to compare our BDEC predicted parcellation with nine other parcellations, Figure 7 and Table 1 shows that our BDEC parcellations represent excellent parcel homogeneity. Specifically, from Table 1, our BDEC-removed has significantly higher regional homogeneity than all other parcellations, with a z-score of 10.41, which is 591% higher than that of the Schaefer parcellation with the highest homogeneity. There are two reasons for this significant difference. First, our model was trained to obtain high homogeneity parcellations, so our model parcellations themselves have high homogeneity. Second, if a parcellation is not completely connected in space, it is more likely to obtain a larger z-score value (Arslan et al., 2018). Furthermore, we compared the BDEC-merged, which is completely connected spatially, with other parcellations and found that its z-score value is also higher than other parcellations, and its homogeneity is 124% higher than that of the Schaefer parcellation. In addition, we set up different post-processing options to prove that they have less impact on the result (**see Supplementary Table S2**) and use another data set (Kliemann et al., 2019) to show that BDEC parcellations have strong generalization (**see Supplementary Figure S1**).

One notable result is that the inventor of the null model, Gordon, performs the worst in our experiment. We speculate that this may because: 1) the dilated Gordon parcellation is used in this experiment, which may reduce its homogeneity; 2) the fMRI data used in our experiment is different from that used by (Gordon et al., 2016).

### 3.4. Parcel Homogeneity with DCBC

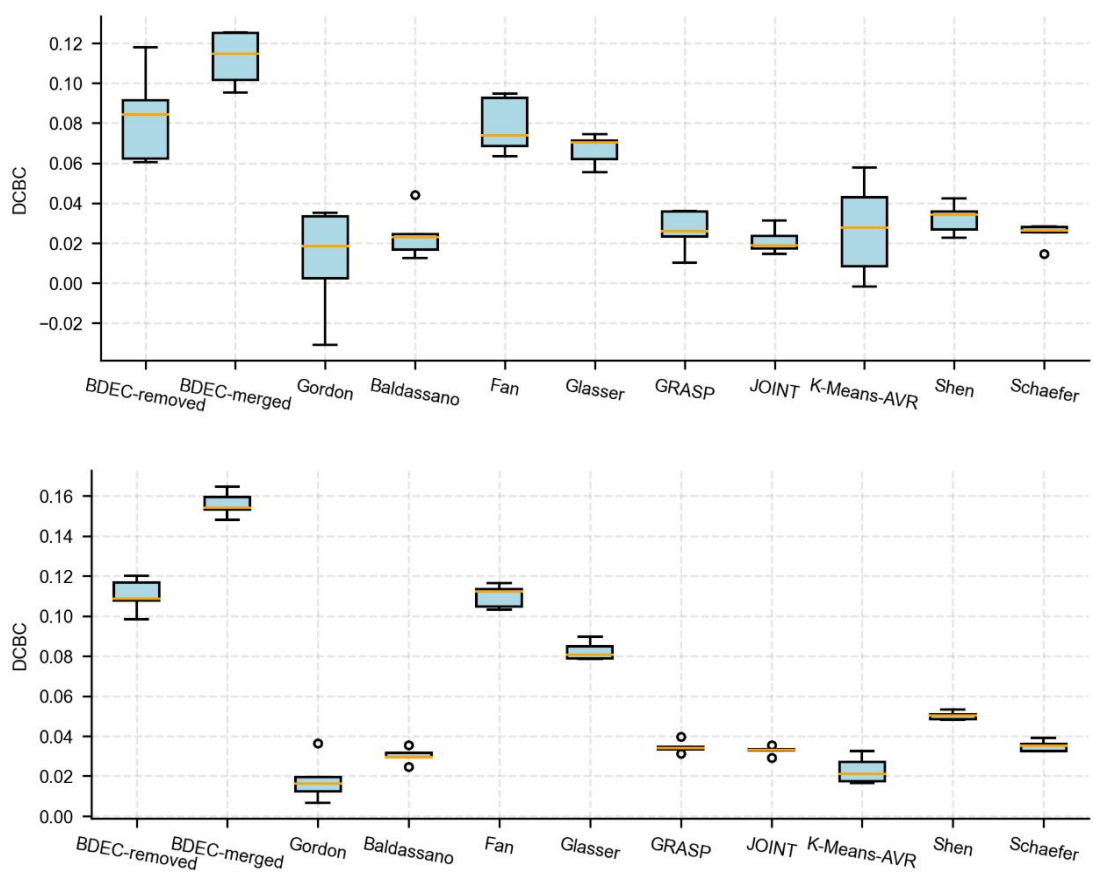

**Figure 8.** DCBC values, compared within the selected parcellations. Also use five-fold cross-validation on the whole test set, **Top:** using the average data of validate set, the mean DCBC values (yellow line) of our BDEC parcellations are highest. **Bottom:** using the average data of sub-test set, the mean DCBC value of BDEC-merged is highest and BDEC-removed is third (Fan is the Second).

**Table 2.** It shows the DCBC values, using average data of the full test set.

| Parcellation | DCBC value |
| --- | --- |
| **BDEC-merged** | **0.1579** |
| Fan | 0.1116 |
| **BDEC-removed** | **0.1110** |
| Glasser | 0.0825 |
| Shen | 0.0516 |
| Schaefer400 | 0.0361 |
| GRASP | 0.0357 |
| JOINT | 0.0343 |
| Baldassano | 0.0307 |
| K-Means-AVR | 0.0213 |
| Gordon | 0.0179 |

Under the DCBC method, Table 2 demonstrates that our BDEC predicted parcellation has higher functional homogeneity compared to the other 9 parcellations. The DCBC values for the removed and merged parcellations are 0.1110 and 0.1579, respectively, both of which are almost higher than the DCBC values of the rest parcellations. The Fan parcellation has highest DCBC value among the 9 parcellations, which is 0.1116. Figure 8 illustrates that our DBEC parcellations have almost the highest DCBC values on both the small validation set and the larger sub-test set. We also explored for parcel homogeneity of BDEC parcellations using DCBC in individual subjects, which presented comparable results (**see Supplementary Figure S2**).

## 3.5. Network analysis with gender classification

Table 3. The average accuracy of gender prediction was calculated using ten-fold cross-validation on a randomly selected group of 800 individuals (400 males and 400 females) from the HCP S1200 release dataset.

| Parcellation | accuracy |
| --- | --- |
| **BDEC-merged** | **0.7550** |
| **BDEC-removed** | **0.7425** |
| Glasser | 0.7413 |
| GRASP | 0.7388 |
| Baldassano | 0.7375 |
| JOINT | 0.7338 |
| Gordon | 0.7338 |
| K-Means-AVR | 0.7300 |
| Shen | 0.7300 |
| Fan | 0.7275 |
| Schaefer | 0.6737 |

According to Table 3, our BDEC parcellations have higher accuracy in gender classification than the other nine parcellations, regardless of the post-processing method used. The gender prediction accuracies of the remove and merged parcellations are 0.7425 and 0.7550, respectively. Since male and female brains have been found to have significantly different structural and functional connectivity networks, higher gender classification accuracy suggests higher parcellation quality to some extent.

We found an interesting phenomenon that the accuracy of Glasser parcellations is the highest among the nine parcellations and very close to our model parcellations' accuracy. One possible explanation is that Glasser parcellations use a multimodal approach, incorporating information on neurobiology, and therefore have a high accuracy in gender prediction.

## 3.6. Parcel validity with Silhouette coefficient

**Table 4.** The Silhouette coefficients of each parcellation are calculated, using the average fMRI data of the full test set. The results are sorted in descending order according to the Silhouette coefficient.

| Parcellation | accuracy |
| --- | --- |
| **BDEC-removed** | **0.3559** |
| **BDEC-merged** | **0.3356** |
| Fan | 0.3215 |
| Gordon | 0.3138 |
| Schaefer400 | 0.3127 |
| JOINT | 0.3051 |
| K-Means-AVR | 0.3043 |
| GRASP | 0.2941 |
| Shen | 0.2910 |
| Glasser | 0.2835 |
| Baldassano | 0.2703 |

Table 4 shows that our two BDEC parcellations both have higher Silhouette coefficients than the other parcellations, which indicates higher quality of our parcellations to some extent. The Silhouette coefficient combines the similarity among samples within clusters and the dissimilarity among samples across clusters. From this perspective, a higher Silhouette coefficient indicates that our parcellations are more reasonable.

## 3.7. Task Homogeneity with null model

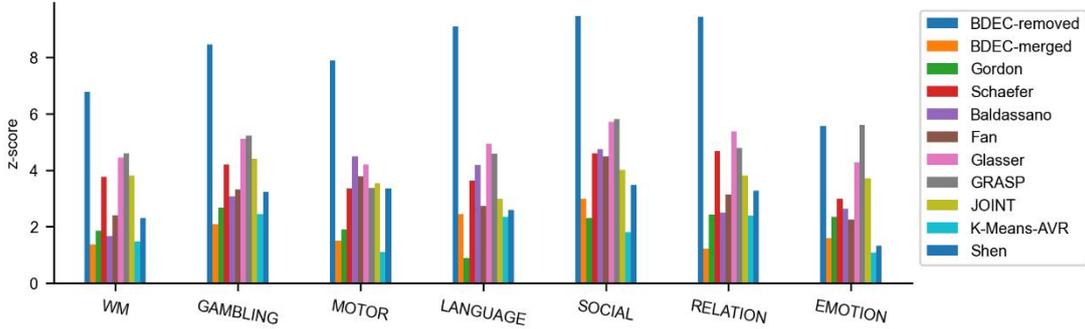

**Figure 9.** Compute the task activation variance for the 7 tasks using the null model approach.

Figure 9 shows that BDEC-removed parcellation performs better than other parcellations in all tasks. It is worth noting that the BDEC-merged parcellation is not the best, but it is not the worst either. We speculate that the merging step used in post-processing method may have influenced the final results since it was not applied to other parcellations.

Overall, both of our BDEC parcellation methods present a considerable performance under all tasks, especially under the 'LANGUAGE' and 'SOCIAL' tasks, which could provide some guidance for brain network analysis tasks based on these two tasks.

## 4. Discussion

In this paper, we propose a deep learning-based parcellation method, BDEC, without introducing model biases induced by the assumptions of employed parcellation models. We treat parcellation as a deep clustering issue and consider both functional activation patterns of brain dynamics and the locations of cortex structures. That is, we incorporate signal homogeneity within parcels, signal heterogeneity among parcels and the location embedding of each parcel into our algorithmic design. Specifically, we first utilize the encoder-decoder framework to obtain the latent representation of functional activation. Second, we perform soft embedding on the location information and sum it to the representation of functional activation. Then, a particular distribution is designed and used to project the distribution of the samples on it to minimize the within-cluster distance. Between-cluster distance is denoted by the distance among cluster centers. After incorporating such information into our loss function, we perform the rs-fMRI-based parcellation through training the deep clustering algorithm. Compared with previous work, our method benefits from its assumption-free characteristics and demonstrates the effectiveness of using deep learning techniques in this field. Our parcellation presents an improved homogeneity within parcels, validity, correspondence of cognitive tasks and demonstrated performance on the network analysis. And the results present a relatively robust performance on another dataset.

### 4.1. BDEC Parcellations Compare with Other Parcellations

We perform extensive comparisons with other methods. We selected four categories of brain parcellations: the first category is based on traditional clustering algorithms (Shen; JOINT), which typically focus on global signal similarity but overlook spatial connectivity or tend to generate parcels of similar sizes; The second category is based on region-growing algorithms (Gordon), which emphasize local signal variations but are sensitive to the initial seed selection; The third category considers both global signal similarity and local variations (Schaefer; Baldassano; GRASP; K-Means-AVG), but these parcellations are prone to introduce model biases; The finally category combines anatomical information of the brain (Glasser; Fan), going beyond purely data-driven parcellations. These four categories roughly cover existing parcellation types, and we select nine representative parcellations from them to ensure a more comprehensive and reasonable comparative analysis. Overall, our BDEC parcellations demonstrate nearly the highest level of parcel homogeneity compared to the other nine parcellations, and a clearer boundary among parcels. This high degree of homogeneity in BDEC indicates that most parcels represent regions with uniform BOLD signal, which is an expected characteristic of most cortical areas.

However, when comparing the parcel homogeneity across different parcellations, the size and number of parcels significantly impact the results. Given this fact, we selected two metrics, named "null model" which enable homogeneity comparisons to be performed at the same size and number of parcels almost entirely and "DCBC" which are insensitive to parcel size and number. Furthermore, the number of the remaining nine parcellations we chose is roughly similar to the BDEC parcellations

(half of the parcellation whose parcel size being approximately 400). In comparison to directly calculating the Pearson correlation coefficient, we ensure fairness in the comparative analysis through the dual assurance of metrics and the selection of participating comparable parcellations.

It is frustrating that, due to the combined influence of the physiological nature of fMRI signals and data smoothing, there is currently no definitive method that guarantees absolute fairness in the comparison(Gordon et al., 2016). Therefore, finding a perfect metric for parcel homogeneity is challenging.

### 4.2. The Post-Process of BDEC

We perform a soft manner of incorporating spatial information, which follows the common paradigm of previous studies(Arslan et al., 2015; Glasser et al., 2016; Gordon et al., 2016). We found that although we incorporated spatial information into the BDEC model, the constraint is relatively loose, resulting in spatially dispersed parcels. Therefore, some post-processing strategies that can further guarantee the spatial congruity within parcels are introduced in our study. First, we set a threshold called Remove Num (RN). We remove parcels with vertex count smaller than RN and then perform dilation to fill the holes, referred to as BDEC-removed parcellation. Subsequently, we merge the most homogeneous neighboring regions to obtain spatially fully connected parcels, referred to as BDEC-merged parcellation. These two types of parcellations obtained from different post-processing steps possess different characteristics. BDEC-removed exhibits higher parcel homogeneity, while BDEC-merged is more suitable as a template for brain network analysis. The phenomenon suggest that such post-processing technique involved in the parcellation should be selected and further developed in a task-specific manner.

Additionally, we explored the impact of different RN values on parcel homogeneity. As shown in **Supplementary Table S1**, under the premise that the number of parcels does not significantly decrease with post-processing, our BDEC parcellations are not sensitive to the choice of RN. Regardless of the RN value, the parcel homogeneity of BDEC parcellations is consistently higher than the other nine parcellations. This phenomenon indicates that the BDEC model indeed captures functional patterns of the brain and is insensitive to specific post-processing steps.

**4.3. BDEC Parcellation Demonstrates Good Generalization Capability**

To assess the generalization capability of the BDEC parcellation, we utilized a six-subject dataset obtained from OpenNeuro. We test the generalization ability use the parcellation of the model trained by the HCP dataset.

As shown in **Supplementary Figure S1**, we compared the parcel homogeneity (using null model) and validity (using silhouette coefficient) of the BDEC parcellations with the remaining nine parcellations. Although the BDEC parcellations did not consistently yield the best results, they outperformed the majority of the parcellations. The reason behind the BDEC parcellations not being the best in terms of results could be attributed to the difference of the data distribution between the large-scale dataset for training and the much smaller size of the OpenNeuro dataset. BDEC was trained on a large amount of averaged rs-fMRI data, which effectively captured group-level functional features. However, the six-subject dataset is relatively small and exhibited strong individual biases in the averaged data. Despite this, the BDEC parcellations still showed favorable experimental results, demonstrating their good generalization capability across datasets.

**4.4. Why Is Deep Learning Rarely Used for Resting-State Functional Parcellation?**

Deep learning shows powerful data fitting capabilities, which gives the BDEC parcellations an unparalleled advantage in terms of parcel homogeneity. However, we have not found any relevant work applying deep learning to unsupervised parcellation of rs-fMRI data. Instead, we have come across studies that have used it for supervised parcellation using MRI(Zhao et al., 2021b, 2021a, 2019). Cortical parcellation is fundamentally a deep clustering problem, and there have been well-established studies in image clustering (Dizaji et al., 2017; Guo et al., 2017). So, what could be the reasons for the occurrence of this peculiar phenomenon? We analyze it from two perspectives. First, there is a significant difference between image data and rs-fMRI data. Image data usually requires minimal preprocessing, while rs-fMRI data necessitates complex preprocessing and contains different information. Consequently, the issue of clustering degradation, where the final number of clusters is much smaller than the specified quantity, becomes more severe in rs-fMRI data. Second, cortical parcellation expects the parcels to be spatially connected to achieve stronger physiological interpretability. However, incorporating corresponding constraints in deep learning is challenging.

Despite these challenges, we have high expectations for the application of deep learning in cortical parcellation and look forward to further advancements in this field.

**4.5. Limitations and Future Work**

This work focuses solely on the parcellation of the cerebral cortex. The brain is a complex organ consisting of both cortical and subcortical structures that are interconnected in space. However, our parcellation approach only targets the cortical regions, which may limit our understanding of brain functional organization and connectivity. Nevertheless, our deep clustering method can also be applied to subcortical structures. However, due to significant signal-to-noise ratio differences between the cortical and subcortical regions, achieving accurate whole-brain parcellation using a single model is challenging (Schaefer et al., 2018). Therefore, studying the cortical and subcortical regions separately is a reasonable choice.

The BDEC parcellation does not align well with anatomical boundaries. BDEC cortical parcellation, being driven solely by functional characteristics, is a pure functional parcellation. While it considers spatial connectivity, the complex multi-scale functionality of certain cortical areas makes it difficult to directly obtain their anatomical boundaries from rs-fMRI data. We use the Dice coefficient to measure the similarity between BDEC parcellations and the anatomical AAL parcellation, with a range of values from 0 to 1. The Dice coefficient values for BDEC-removed and BDEC-merged are 0.06 and 0.07, respectively. However, this does not necessarily indicate lower quality of the parcellation. (Felleman and Van Essen, 1991) emphasized the significance of utilizing multiple modalities to uniquely identify each cortical area in their influential study on macaque visual cortex parcellation. These modalities encompass connectivity, architectural features, topographic organization, functional responses, and lesion-induced behavioral consequences. However, their findings indicated that not all approaches were effective in identifying all cortical areas. In many cases, only one or two modalities were able to differentiate specific areas, suggesting that a comprehensive categorization of the human cortex would require supplementary data from additional modalities.

The BDEC parcellations are not applicable to individual subjects. We calculated the parcel homogeneity of different parcellations on a dataset of 100 individual subjects, but the results are noticeably lower compared to the group level and exhibited high variance (**see Supplementary Figure S2**). Since our cortical parcellation is derived from averaging multiple subject data, it represents a group-level parcellation. It is well-known that individual brains exhibit variability in functional organization(Bergmann et al., 2020; Glasser et al., 2016; Gordon et al., 2016; Laumann et al., 2015), and brain function can differ among individuals due to factors such as gender, age, and disease status. Therefore, a group-level parcellation template may not capture individual functional differences, leading to inconsistent functional outcomes in the parcellation results. This issue may be exacerbated when using higher-resolution parcellation templates, as fine-grained structural differences between individuals become more pronounced. Our future work involves incorporating additional spatial constraints into the model and increasing model complexity to extend the approach to individual subjects.

5. **Conclusions**

This study demonstrates the feasibility of using deep learning for functional parcellation of human cerebral cortex based on rs-fMRI signal. Our model draws on the idea of image deep clustering and is optimized for group level cortical functional parcellation task. By comparing with nine commonly used brain parcellation methods on two separate datasets, our model demonstrates significantly improved functional homogeneity on multiple indicators. Furthermore, it exhibits favorable results in terms of validity, network analysis, task homogeneity, and generalization capability. This work would aid the development of involving deep learning techniques into understanding the organization of brain functional measures.

**Ethical statement**

**Acknowledgements**

## Credit authorship contribution statement

**Xiaoxiao Ma:** Conceptualization, Software, Validation, Data curation, Analysis, Writing – original draft, Writing - review & editing, Project administration

**Chunzhi Yi:** Resources, Conceptualization, Software, Writing - review & editing

**Zhicai Zhong:** Software, Data curation, Writing - review & editing

**Hui Zhou:** Analysis, Writing - review & editing

**Baichun Wei:** Project administration, Writing - review & editing

**Haiqi Zhu:** Validation, Writing - review & editing

**Feng Jiang:** Resources, Writing - review & editing

## Data availability

The Human Connectome Project datasets can be obtained at www.humanconnectome.org/ after presenting request describing the intended use of the data.

The Caltech rsfMRI Dataset used in supplementary materials can be obtained at https://openneuro.org/datasets/ds002232/versions/1.0.0.

The group average data of the above two datasets is available upon request to Xiaoxiao Ma (jmgytc@163.com).

## Code availability

The BDEC model and parcellations are submitted in https://github.com/jmgytc/BDEC, implemented with Python.

**Conflict of interest**

The authors declare that they have no known competing financial interests or personal relationships that could have appeared to influence the work reported in this paper.

# Appendix. Supplementary materials

**Table S1: All parcellations represented in this work are shown.** Except for the two BDEC parcellations, the other nine parcellations can be found in [Brain Parcellation Survey – BioMedIA (ic.ac.uk).](Brain Parcellation Survey – BioMedIA (ic.ac.uk).)

| Parcellation | Resolution | Description |
| --- | --- | --- |
| BDEC-removed | 394(198L,196R) | The parcellation generated by our BDEC model are lightly post-processed. It removes fewer than 20/9 vertices on left/right hemisphere separately and then expands to fill the void |
| BDEC-merged | 394(198L,196R) | The parcellation generated by our BDEC model are post-processed. It merges the area with the smallest area into the nearby area, ensuring the spatial connectivity of the parcellation |
| Schaefer | 400(200L,200R) | It is generated by gradient-weighted Markov random fields. This method not only guarantees the global similarity, but also makes the boundary of the region parcels clearer. We chose a parcellation with a resolution of 400 to compare with our parcellation |
| Baldassano | 171(84L,84R) | It is generated by a multi-purpose parameter-free Bayesian clustering model and acts with a dense connection matrix |
| JOINT | 400(200L,200R) | A surface-based parcellation method based on a joint spectral decomposition of individual subjects |
| K-Means-AVR | 400(200L,200R) | The k-means algorithm is used for group average matrix clustering which concatenates spatial coordinates. This approach improves spatial connectivity |
| Fan | 210(105L,105R) | A volumetric brain parcellation is obtained using both anatomical landmarks and connectivity-driven |
| | | information |
| GRASP | 404(202L,202R) | Using Markov model, by adding shape a priori. It can produce spatially connected parcellation |
| Shen | 200(102L,98R) | A spectral clustering approach is used to compute a volumetric groupwise parcellation |
| Glasser | 360(180L, 180R) | A semi-automatic method was used to generate cortical parcellation from multimodal images of 210 adult from HCP |
| Gordon | 333(161L,172R) | A surface-based parcellation calculated from the mean gradient of a functional resting state connected network. This parcellation is iteratively dilated until it covers the entire surface |

**Table S2: DCBC values of whole brain, using average data of the full test set of HCP.** The sign "*" in BDEC-removed-* and BDEC-merged-* represent the vertex remove num (RN) and parcels whose size less than it will be removed. When RN is 25, a large portion of parcels will be removed, so the value of BDEC become low. For the sake of result readability, we only varied the RN value in the left hemisphere, while keeping the RN value in the right hemisphere constant at 9.

| Parcellation | DCBC value |
| --- | --- |
| BDEC-removed-0 | 0.1242 |
| BDEC-removed-5 | 0.1232 |
| BDEC-removed-10 | 0.1185 |
| BDEC-removed-15 | 0.1163 |
| BDEC-removed-20 | 0.1110 |
| BDEC-removed-25 | 0.0666 |
| BDEC-merged-0 | 0.0962 |
| BDEC-merged-5 | 0.1662 |
| BDEC-merged-10 | 0.1537 |
| BDEC-merged-15 | 0.1737 |
| BDEC-merged-20 | 0.1579 |
| BDEC-merged-25 | 0.1022 |
| Fan | 0.1116 |
| Glasser | 0.0825 |
| Shen | 0.0516 |
| Schaefer | 0.0361 |
| GRASP | 0.0357 |
| JOINT | 0.0343 |
| Baldassano | 0.0307 |
| K-Means-AVR | 0.0213 |
| Gordon | 0.0179 |

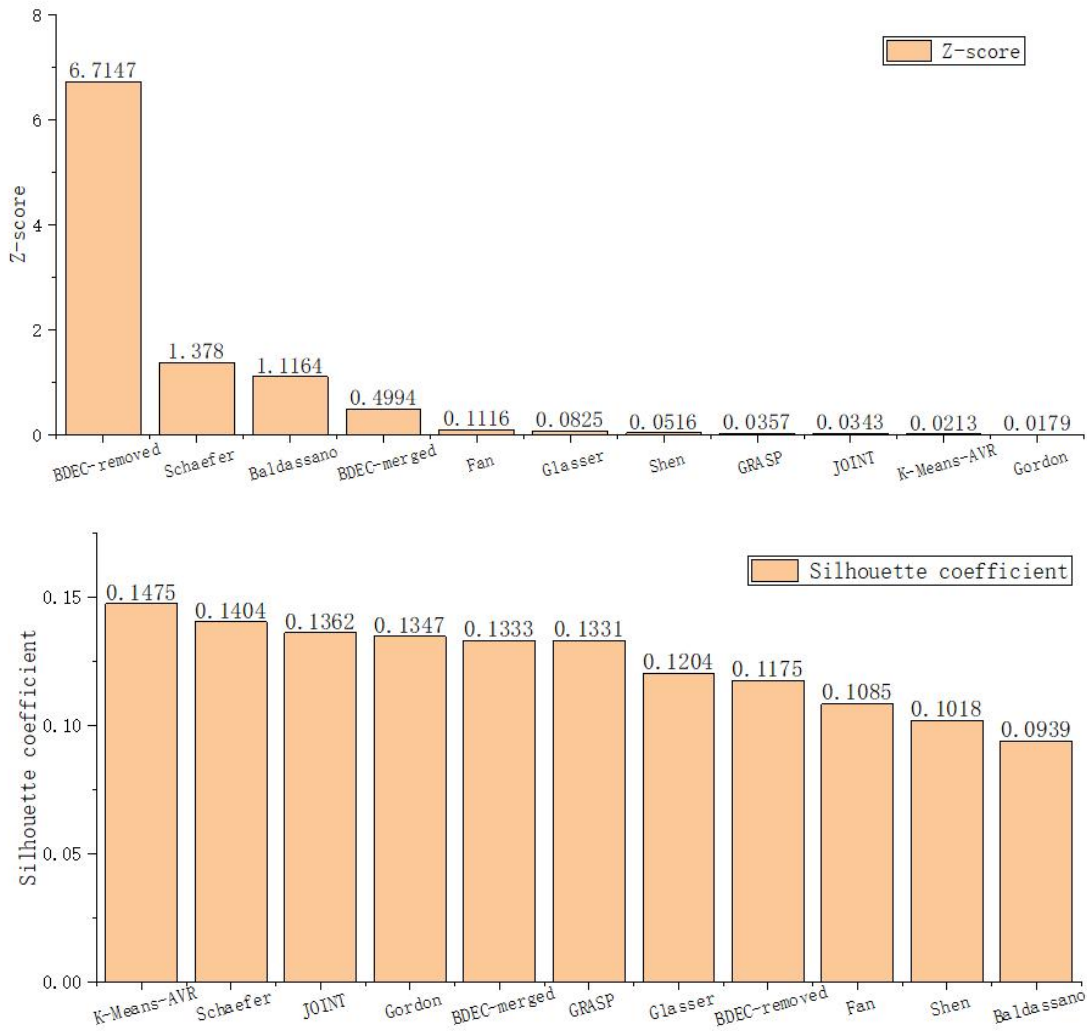

**Figure S1**：The z-score and Silhouette coefficient, using average data of Caltech rs-fMRI Dataset in OpenNeuro which consists of six people and be pre-process by fmriprep.

**Addition**：Caltech rsfMRI Dataset for "Intrinsic functional connectivity of the brain in adults with a single cerebral hemisphere" by Dorit Kliemann, Ralph Adolphs, J. Michael Tyszka, Bruce Fischl, B.T. Thomas Yeo, Remya Nair, Julien Dubois, Lynn K. Paul. Subsequently, the data were preprocessed using the default pipeline of fmriprep(v-23.0.2) and projected onto the fs_LR_32k surface space. The datasets were then averaged to obtain group-level rs-fMRI data.

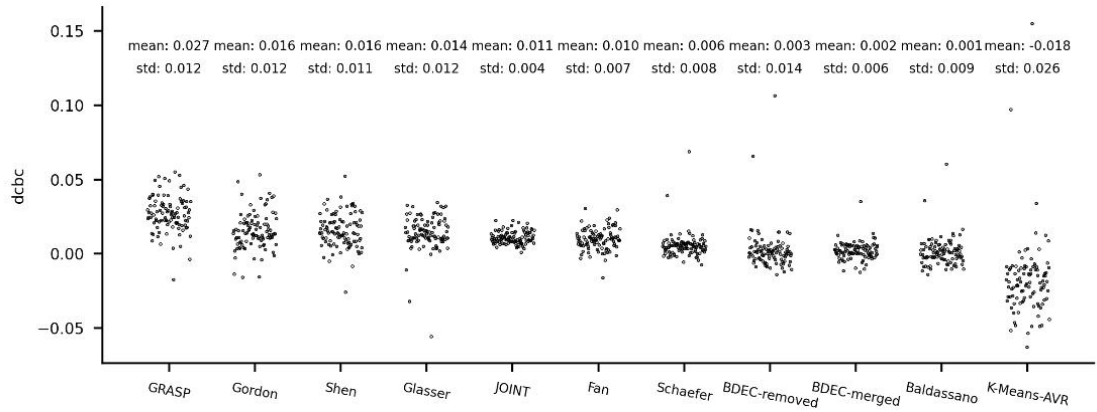

**Figure S2：Randomly selecting 100 participants from the full test dataset of HCP, we calculated their DCBC values.** This was done to assess the performance of the group-level parcellation template on individual participants.